\documentclass[11pt,preprint]{emulateapj}

\newcommand{\kms}{\,km~s$^{-1}$}

\newcommand{\etal}{{et al.~}}
\newcommand{\lta}{\lesssim}

\newcommand{\Msun}{\>{\rm M_{\odot}}}

\def\spose#1{\hbox to 0pt{#1\hss}}
\def\simlt{\mathrel{\spose{\lower 3pt\hbox{$\mathchar"218$}}
     \raise 2.0pt\hbox{$\mathchar"13C$}}}
\def\simgt{\mathrel{\spose{\lower 3pt\hbox{$\mathchar"218$}}
     \raise 2.0pt\hbox{$\mathchar"13E$}}}
\font\smcap=cmcsc10

\shorttitle{Local Group dE Dynamics}
\shortauthors{Geha~et~al.}

\begin{document}


\title{Local Group Dwarf Elliptical Galaxies:  II.\ 
Stellar Kinematics to Large Radii  in NGC~147 and NGC~185}


\author{M.\ Geha\altaffilmark{1}}

\author{R.\ P.\ van der Marel\altaffilmark{2}} 

\author{P.\ Guhathakurta\altaffilmark{3}}

\author{K.\ M.\ Gilbert\altaffilmark{4}}

\author{J.\ Kalirai\altaffilmark{2}}

\author{E.\ N.\ Kirby\altaffilmark{5}}

\altaffiltext{1}{Astronomy
  Department, Yale University, New Haven, CT~06520.
  marla.geha@yale.edu}

\altaffiltext{2}{Space Telescope
    Science Institute, 3700 San Martin Drive, Baltimore, MD~21218}

\altaffiltext{3}{UCO/Lick Observatory, University of California,
   Santa Cruz, 1156 High Street, Santa Cruz, CA~95064.}

 \altaffiltext{4}{Department of Astronomy, University of Washington,
   Seattle, WA 98195-1580}

 \altaffiltext{5}{Hubble Fellow, California Institute of Technology, Department of Astronomy, 1200 E California Blvd, Pasadena, CA 91125}


\submitted{ }

\begin{abstract}
\renewcommand{\thefootnote}{\fnsymbol{footnote}}

We present kinematic and metallicity profiles for the M\,31 dwarf elliptical (dE) satellite galaxies NGC~147 and NGC~185.  The profiles represent the most extensive spectroscopic radial coverage for any dE galaxy, extending to a projected distance of eight half-light radii ($8r_{\rm eff} \sim 14'$).  We achieve this coverage via Keck/DEIMOS multislit spectroscopic observations of 520 and 442~member red giant branch stars in NGC~147 and NGC~185, respectively.  In contrast to previous studies, we find that both dEs have significant internal rotation.  We measure a maximum rotational velocity of $17\pm 2$\kms\ for NGC~147 and $15\pm5$\kms\ for NGC~185.  While both rotation profiles suggest a flattening in the outer regions, there is no indication that we have reached the radius of maximum rotation velocity.  The velocity dispersions decrease gently with radius with an average dispersion of $16\pm 1$\kms\ for NGC~147 and $24\pm1$\kms\ for NGC~185.  The average metallicity for NGC~147 is [Fe/H] = $-1.1\pm0.1$ and for NGC~185 is [Fe/H] = $-1.3\pm0.1$; both dEs have internal metallicity dispersions of 0.5\,dex, but show no evidence for a radial metallicity gradient.  We construct two-integral axisymmetric dynamical models and find that the observed kinematical profiles cannot be explained without modest amounts of non-baryonic dark matter.  We measure central mass-to-light ratios of $M/L_V = 4.2\pm0.6$ and $M/L_V = 4.6\pm0.6$ for NGC~147 and NGC~185, respectively.   Both dE galaxies are consistent with being primarily flattened by their rotational motions, although some anisotropic velocity dispersion is needed to fully explain their observed shapes.  The velocity profiles of all three Local Group dEs (NGC~147, NGC~185 and NGC~205) suggest that rotation is more prevalent in the dE galaxy class than previously assumed, but is often manifest only at several times the effective radius.  Since all dEs outside the Local Group have been probed to only inside the effective radius, this opens the door for formation mechanisms in which dEs are transformed or stripped versions of gas-rich rotating progenitor galaxies.


\end{abstract}




\section{Introduction}\label{intro_sec}
\renewcommand{\thefootnote}{\fnsymbol{footnote}}

Dwarf elliptical (dE) galaxies are characterized by low surface
brightness ($\mu_{\rm eff,V} > 22$ mag arcsec$^{-2}$), little to no
gas, and old to intermediate age stellar populations.  These galaxies are
spatially clustered and account for more than 75\% of objects brighter
than $M_V < -14$ in nearby galaxy clusters
\citep{ferguson91a,ferguson94a}.  Since there are few, if any,
isolated dE galaxies, most proposed dE galaxy formation scenarios are 
environmentally-driven, such as via gas stripping or gravitational
harassment of gas-rich spiral or dwarf irregular galaxies
\citep[e.g.\,][]{dekel86a, moore98a,Aguerri09a}.  Recent evidence for
embedded stellar disks and bars in some dEs support these scenarios
\citep{jerjen00a,geha05a,lisker06a,chilingarian07a}. Since gas-rich systems
have significant rotation velocities, a key test of this hypothesis
class is comparing the kinematics of dE and gas-rich galaxies.

Internal kinematic measurements for dEs in the
Fornax and Virgo Cluster have revealed an unexpected dichotomy:
roughly half of the observed dEs have significant rotation about the
major-axis, while the remaining galaxies show no detectable major-axis
rotation \citep{pedraz02a,geha03a, vanzee04a, deRijcke05a}.  dE galaxies with no
rotation present a challenge to any scenario in which dE progenitors
are gas-rich and thus rotating \citep{moore98a}.  Ram pressure stripping
alone does not affect the kinematics of the collisionless stellar
component.  Gravitational processes such as galaxy harassment can
reduce the amount of rotational angular momentum, converting this into
random velocity dispersion, however, simulations are unable to
completely remove rotational support \citep{mayer01a,mastropietro05a}.

Because dEs have low surface brightnesses, integrated light
(long-slit) kinematic observations of dEs in both galaxy clusters and
the Local Group are currently limited to within the half-light
effective radius ($r_{\rm eff}$).  Thus, it is possible that
significant rotational angular momentum lies at larger distances than
has so far been explored.  Addressing this question requires
kinematics tracers at larger radius.  While the dynamics of dE globular
cluster systems suggest that there may be significant rotation at
large radius \citep{beasley06a,beasley09a}, the small number of
clusters limits the usefulness of this tracer. The dEs in the Local
Group share the same general properties as the more numerous dEs in galaxy
clusters, however, their proximity allows us to resolve individual
stars \citep{geha06a}, and thus trace kinematics profiles down to arbitrarily
low surface brightness and large radius.

The Local Group contains three dE galaxies, NGC~205, NGC~147 and NGC~185, which in addition to the compact elliptical M32, are the brightest satellites around the spiral galaxy M\,31.   We follow  the terminology conventions of,  e.g.~\citet{bender92a}, although alternatives are in use. While the differing terminologies can be confusing, the key point is that NGC~205, NGC~147, and NGC~185 on the one hand, and M\,32 on the other hand, should not be grouped in a single class. The former three galaxies have fundamental plane properties that indicate similarity to galaxies in the dwarf spheroidal class (dSph). By contrast, galaxies such as M\,32 have properties that indicate similarity to (giant) elliptical galaxies. This likely indicates a difference in formation history, as discussed more fully in, e.g., \citet{kormendy09a}.

The prototype of the dE galaxy class is NGC~205 which is in very close projection to M\,31 ($40' = 8$\,kpc).  Both photometric and kinematic evidence suggests that NGC~205 is tidally interacting with M\,31 \citep{choi02a,geha06a}, and may in fact be on its first orbital approach \citep{howley08a}.  In contrast, NGC~147 and NGC~185 lie at a projected distance of $7^{\circ}$\,($\sim$\,$150$\,kpc) from M\,31.   While the metal-poor stellar halo of M\,31 extends to this distance \citep{guhathakurta05a,kalirai06a},  its gravitational influence of M\,31 is not expected to influence the kinematics of these dE satellites (see \S\,\ref{subsec_bound}).    NGC~147 and NGC~185 are separated from each other by merely $58'$, leading to speculation that they may be a bound galaxy pair \citep{baade44a, vandenbergh98a}.   An accurate distance estimate suggests that these two galaxies are physically separated by over 60\,kpc \citep{mcconnachie05a} and may not be bound (however, see \S\,\ref{subsec_bound}).  The large observed distances between these two dEs and their parent M\,31, combined with their relative proximity to the Sun, make NGC~147 and NGC~185 the best available targets for studying the properties of dE galaxies to large radius.

NGC~147 and NGC~185 share several similar fundamental properties, such as absolute luminosity ($M_V\sim -15.5$), half-light radius ($r_{\rm eff} \sim 2' = 0.3$\,kpc), and central velocity dispersion (for exact values see Table~1).  However, they differ markedly in many aspects.  NGC~185 contains some gas, dust and evidence for recent star formation confined to its center \citep{martinez-delgado99a}, while NGC~147 is devoid of gas or dust and shows no sign of recent star formation activity \citep{young97a, sage98a}.  An average metallicity of [Fe/H] = $-1.4$ is inferred for NGC~185 \citep{martinez-delgado99a}, while a more metal rich population of [Fe/H] = $-1.0$ is inferred for NGC~147 \citep{han97a, davidge05a, goncalves07a}.  HST/WFPC2 imaging of both galaxies also implies the presence of intermediate-age stars, with NGC~147 having a more significant contribution than NGC~185 \citep{butler05a}.  Thus, while both dEs are dominated by old to intermediate-age stars, the metallicity and age mixture of these components are different in each galaxy.

The first spectroscopic long-slit observations of Local Group dEs by \citet{bender91a} suggested that while NGC~185 appeared supported entirely by random motions, NGC~147 had a significant rotational component of $6.5\pm1.1$\kms.  Somewhat deeper observations by \citet{simien02a} extended the kinematic profiles out $\sim2'$ (1$r_{\rm eff}$), but did not measure a significant rotation for either galaxy.  In contrast, both studies measured a rotational component in NGC~205 (within the tidal radius), later confirmed by \citet{geha06a} to be $11\pm5$\kms.  Using the Simien \& Prugniel data, \citet{derijcke06a} constructed dynamical models for all three Local Group dEs, concluding that, within the half-light radii, all three galaxies have mass-to-light ratios above that expected from the stellar populations alone.

In this paper, we present accurate radial velocities out to 8\,$r_{\rm eff}$ in the Local Group dE galaxies NGC~147 and NGC~185, based on Keck/DEIMOS multi-slit spectroscopy for 520 and 442~red giant branch (RGB) stars, respectively.  The paper is organized as follows: in \S\,\ref{sec_data} we discuss target selection for our DEIMOS slitmasks, the observing procedure and data reduction.  In \S\,\ref{sec_results}, we discuss the velocity, velocity dispersion and metallicity profiles.  In \S\,\ref{sec_model}, we construct two-integral dynamical models for these galaxies and discuss the best fit mass distribution and dark matter content.  Finally, in \S\,\ref{sec_disc}, we discuss the implications of these results for the formation of dE galaxies.

Throughout this paper we adopt distance moduli determined by \citet{mcconnachie05a} via the tip of the RGB method.  The distance modulus for NGC~147 is $(m -M)_0 = 24.43 \pm 0.04$ ($675\pm 27$\,kpc) and NGC~185 is $(m -M)_0 = 24.23 \pm 0.03$ ($616\pm 26$\,kpc).  This places NGC~147 and NGC~185 at projected distances of 140 and 185\,kpc from their parent galaxy M\,31, respectively.


\begin{figure*}[t!]
\plotone{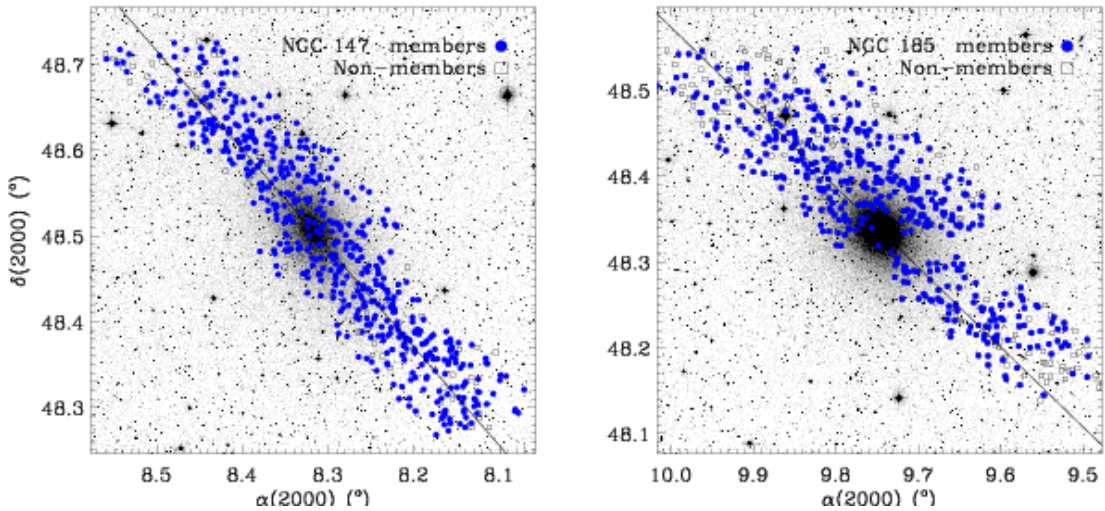}
\caption{Palomar Sky Survey images of the two Local Group dE galaxies,
 NGC~147 ({\it left}) and NGC~185 ({\it right}). Images are
  $30'\times 30'$; North is up, East is to the left. The major-axis of
  each galaxy is indicated by the solid line.  The spatial position of
  Keck/DEIMOS spectroscopically confirmed member stars are shown as
  blue circles, non-members are plotted as open
  squares.  \label{fig_dss}}
\end{figure*}

\section{Data}\label{sec_data}

\subsection{Target Selection}\label{subsec_targets}

\begin{figure*}[t!]
\plotone{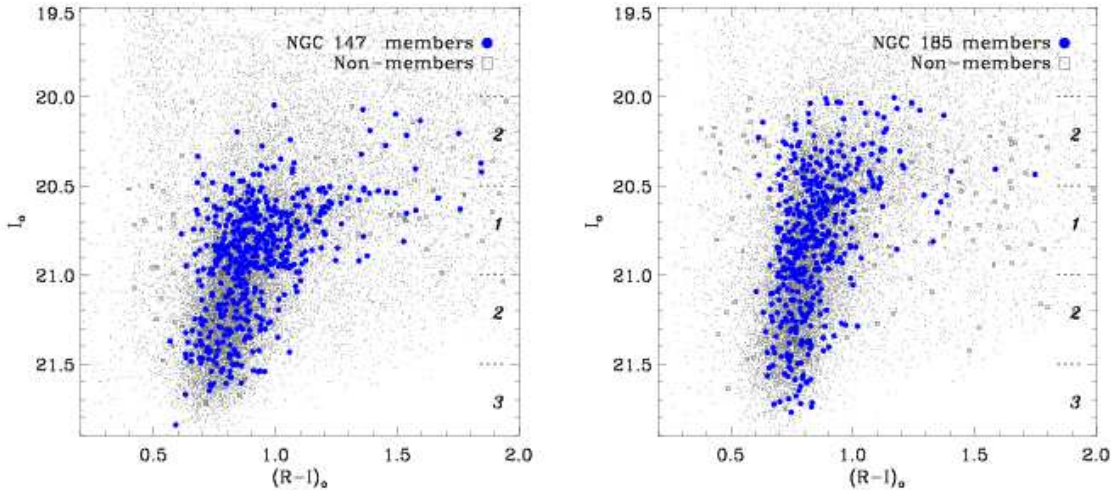}
\caption{Color-magnitude diagrams for NGC~147 ({\it left}) and NGC~185
  ({\it right}) based on \citet{battinelli04a,battinelli04b} $R$- and
  $I$-band CHFT12K photometry, corrected for extinction, in a $42'\times28'$ region centered on  each galaxy.  Blue circles are Keck/DEIMOS spectroscopically
  confirmed members of each galaxy, open squares are non-members.  The
  numbered regions indicate the location of our primary (1), secondary
  (2), and tertiary (3) spectroscopic priorities. \label{fig_cmd}}
\end{figure*}

Individual stars in NGC~147 and NGC~185 were targeted for spectroscopy according to their probability of being a RGB star based on Canada-France-Hawaii Telescope CFH12K mosaic imaging in the $R$- and $I$-bands \citep{battinelli04a,battinelli04b}. The CCD mosaic covers a $42'\times28'$ region centered on each galaxy with $0.206''$ pixels.  Stellar photometry from these images extends 2\,mag below the tip of the RGB.  The spatial distribution of stars is shown in Figure~\ref{fig_dss}.  Color-magnitude diagrams are shown in Figure~\ref{fig_cmd}. 

The absolute magnitude of the RGB for a metal-poor stellar population is $M_I\sim -4$.  At the assumed distance and reddening of NGC~147 and NGC~185, the apparent magnitude of the RGB population is $I_o \sim 20.3$ (see Figure~\ref{fig_cmd}).  Since internal metallicity variations will cause a spread in the colors of RGB stars, we select spectroscopic targets based primarily on apparent magnitude rather than color.  We note there is a 0.1-0.2\,mag photometric zero-point error between individual CCD chips in the CFHT photometry which we discovered after our spectroscopic observations were completed.  While this led us to use an alternative dataset to determine the surface brightness profiles in \S\,\ref{ss:prof}, the error had little to no affect on our targeting efficiencies.

For both dEs, highest priority in the spectroscopic target list was assigned to stars between $20.5 \le I_o \le 21.0$.  Second priority was given to stars between $20.0 \le I_o < 20.5$ and $21.0 < I_o \le 21.5$; lowest priority was assigned to objects $I_o > 21.5$.  To minimize Galactic foreground contamination, targets were required to have $(R-I)_o > 0.2$.  Stars with photometric errors larger than $\sigma_{I} > 0.1$ were rejected, as were stars having neighbors of equal or greater brightness within a radius $r< 4''$.  
The photometric criteria used for spectroscopic target selection are shown in Figure~\ref{fig_cmd}.

Slitmasks were created using the DEIMOS {\tt dsimulator} slitmask
design software.  For each slitmask the software fills in the mask
area to the extent possible with the highest priority input targets.
It then fills in the remaining space on the slitmask with lower
priority targets.  The specifics of each mask are listed in
Table~2.

\subsection{Observations and Data Reduction}\label{subsec_redux}

The data were taken with the Keck~II 10-m telescope and the DEIMOS
spectrograph \citep{faber03a}.  Four multislit masks were observed in
NGC~147 on the nights of August 28 and September 5-7, 2005.  Five
masks were observed in NGC~185 on September 16, 2006.  Mask positions,
exposure times and other observing details are given in Table~2.  The
masks were observed with the 1200~line~mm$^{-1}$\,grating covering a
wavelength region $6400-9100\mbox{\AA}$.  The spectral dispersion of
this setup is $0.33\mbox{\AA}$, and the resulting spectral resolution,
taking into account the anamorphic distortion, is $1.37\mbox{\AA}$
(FWHM, equivalent to 47\kms\ at the Ca II triplet). The spatial scale
is $0.12''$~per pixel and slitlets were $0.7''$ wide.  The minimum
slit length was $4''$ which allows adequate sky subtraction; the
minimum spatial separation between slit ends was $0.4''$ (three
pixels).  

Spectra were reduced using a modified version of the {\tt spec2d} software pipeline (version~1.1.4) developed by the DEEP2 team at the University of California-Berkeley for that survey. A detailed description of the two-dimensional reductions can be found in \citet{simon07a}.  The final one-dimensional spectra are rebinned into logarithmic wavelength bins with 15\,\kms\ per pixel.  An example DEIMOS spectrum taken with the same settings is shown in Figure~4 of \citet{geha06a}.

\subsection{Radial Velocities and Error Estimates}\label{subsec_rvel}

Radial velocities were measured by cross-correlating the observed
science spectra with a series of high signal-to-noise stellar
templates. Stellar templates were
observed with Keck/DEIMOS using the same setup as described in
\S\,\ref{subsec_redux} and cover a wide range of stellar types (F8 to
M8 giants, subgiants and dwarf stars) and metallicities ([Fe/H] =
$-2.12$ to $+0.11$).  We calculate and apply a telluric correction to
each science spectrum by cross correlating a hot stellar template with
the night sky absorption lines following the method in
\citet{sohn06a}.  The telluric correction accounts for the velocity
error due to mis-centering the star within the $0.7''$ slit caused by
small mask rotations or astrometric errors.  We apply both a telluric
and heliocentric correction to all velocities presented in this paper.

We determine the random component of our velocity errors using a Monte
Carlo bootstrap method.  Noise is added to each pixel in the
one-dimensional science spectrum, we then recalculate the velocity and
telluric correction for 1000 noise realizations.  Error bars are
defined as the square root of the variance in the recovered mean
velocity in the Monte Carlo simulations.  The systematic contribution
to the velocity error was determined by \citet{simon07a} to be
2.2\kms\ based on repeated independent measurements of individual
stars.  The systematic error contribution is expected to be constant
as the spectrograph setup and velocity cross-correlation
routines are identical.  We add the random and systematic errors in
quadrature to arrive at the final velocity error for each science
measurement.  Radial velocities were successfully measured for 1227 of
the 1483 extracted spectra across the nine observed DEIMOS masks.  The
majority of spectra for which we could not measure a redshift did not
have sufficient signal-to-noise.  The fitted velocities were visually
inspected to ensure reliability.

\subsection{Membership Criteria}

The expected velocity distributions of NGC~147 and NGC~185
overlap with both the stellar velocity distribution of the M31 halo and
the foreground Milky Way disk and halo.  Thus, a simple velocity cut is
not sufficient to define membership in these dE galaxies.  We determine
membership in NGC~147 and NGC~185 using a modified version of the
probability method of \citet{gilbert06a}.  This method computes a
membership probability for each star based on the assumption that
dE stars have different physical properties than the contaminating
foreground and background populations.

The \citet{gilbert06a} method computes the probability that a given star is either a RGB star at the distance of NGC~147 or NGC~185, or a foreground Milky Way dwarf star based on the star's location in four spectroscopic and photometric diagnostics: line of sight velocity, strength of the surface-gravity sensitive Na {\smcap II} absorption line at $\sim 8190$\mbox{\AA}, position in a color-magnitude diagram, and a comparison of photometric (CMD-based) and spectroscopic (based on the Ca {\smcap II} triplet at $\sim 8500$\mbox{\AA}) metallicity estimates.  A star's position in a diagnostic is compared to probability distribution functions based on appropriate training sets to determine the likelihood the star is a RGB or foreground dwarf star.  The likelihoods from the individual diagnostics are then combined to produce an overall likelihood. 

For the present analysis, several modifications were made to the method published in \citet{gilbert06a}, which was optimized to identify red giant branch stars in the halo of M31.  Several of the diagnostics utilize Johnson-Cousins $V$ and $I$ magnitudes, while the dE photometry was taken in the $R$ and $I$ bands.  The CMD diagnostic and spectroscopic versus photometric metallicity diagnostic were re-derived using \citet{vandenberg06a} isochrones in ($I$, $R-I$)  shifted to the distance of each dE.  To utilize the Na {\smcap II} diagnostic, VandenBerg isochrones were used to define a relation between $R-I$ and $V-I$, which was then used to derive a $V-I$ color for each star.    The velocity distribution of each dE is modeled as a Gaussian, with a constant velocity dispersion and a central velocity that varies with distance along the semi-major axis.  We expect little contamination from M\,31 halo stars at these radii \citep[$<$\,1 star per mask;][]{guhathakurta05a}.

Stars that are more than three times more likely to be a red giant branch star at the distance of the dE than a foreground dwarf star, and which have a spatial position and line-of-sight velocity consistent with that of the dE, are classified as members. Given these membership criteria, we identify 520 and 442 RGB star as members of the dE galaxies NGC~147 and NGC~185, respectively.   Velocities and velocities errors for individual member stars are listed in Tables~6 and 7 in the electronic version of the paper, or on request from the authors.

\begin{figure*}[t!]
\epsscale{1.15}
\plotone{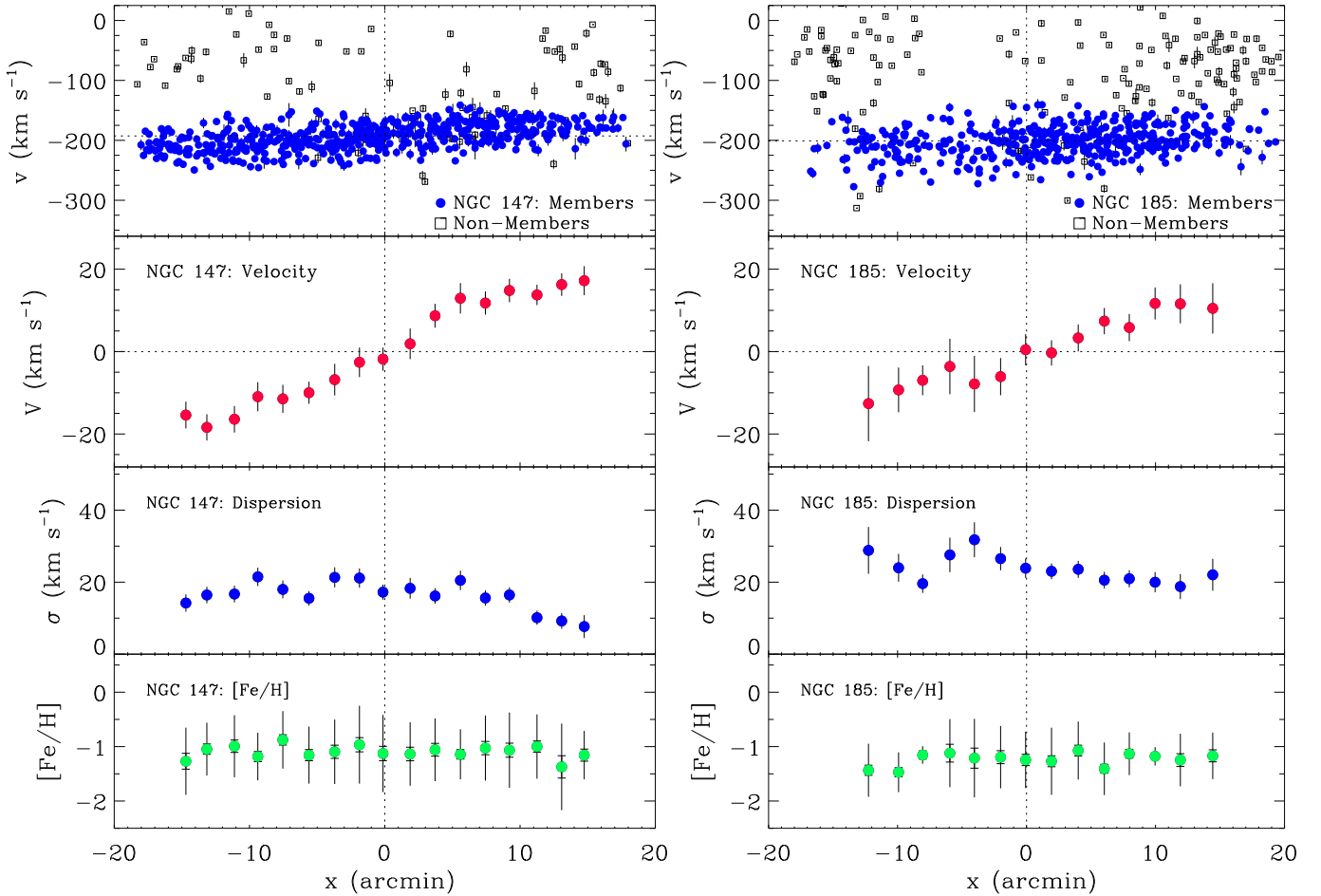}
\vskip 0.25cm
\caption{({\it Top\/}) Individual velocities as a function of semi-major axis distance for dE member stars (blue circles) and non-members (open squares) of NGC~147 ({\it left}) and NGC~185 ({\it right}).  Positive radii correspond to the North-East side of each dE.  ({\it Top middle\/}) The binned major-axis velocity profile showing significant rotation velocities in each dE galaxy. ({\it Bottom middle\/})  The velocity dispersion profiles are nearly flat to the last measured radial point. ({\it Bottom\/})  The [Fe/H] spectroscopic metallicity as a function of semi-major axis distance.  Barred errors indicate the error on the mean metallicity, non-barred errors indicate the metallicity dispersion in each bin.  \label{fig_vp}}
\end{figure*}

\section{Results}\label{sec_results}
 
The measured velocities of individual stars allow us to probe the
dynamics of NGC~147 and NGC~185 to much larger radius than possible
via integrated-light spectroscopy.  In order to compare to previous
results, and to construct dynamical models, we bin the individual
velocity measurements based on their major-axis distance $x$, using
bin widths $\Delta x=2.05'$ for NGC~147 and $\Delta x=1.87'$ for
NGC~185.  The bin widths were chosen to roughly match the effective
radius of each dE.  The bin with integer index number $i$ spans values
of $x$ from $(i-{1\over2}) \Delta x$ to $(i+{1\over2}) \Delta x$. The
value of $x$ for each star was calculated using spherical trigonometry
and a zenithal projection, using the galaxy center positions and major
axis position angles (Table~1 and 3); positive radii lie on the North-East side
of each galaxy.  

We analyze our kinematic data following the procedure described in Appendix~A of \citet{vandermarel09a}.  For each bin number $i$ we calculated the weighted mean velocity $V_i$ of the individual stellar velocities $v_j$, and its formal error $\Delta V_i$. Individual weights were set equal to the quadrature sum of the observational velocity errors $\Delta v_j$ and the velocity dispersion $\sigma_i$ for the given bin. The latter was determined by finding the value that maximizes the likelihood of the set of velocity residuals $v_j-V_i$. The formal error $\Delta \sigma_i$ was calculated using a Monte-Carlo approach, taking into account the observational velocity errors $\Delta v_j$. The systemic radial velocity, $V_{\rm sys}$,  for each galaxy was determined by finding
the value that provides the lowest $\chi^2$ when the dynamical models
described in \S\,\ref{sec_model} were fit to the data and is subtracted from each of the binned major-axis velocities $V_i$.  The systemic velocity for NGC~147 is $V_{\rm sys} = -193.1 \pm 0.8$\kms\ and $V_{\rm sys} = -203.8 \pm 1.1$\kms\ for NGC~185.  This is consistent with literature values from \citet{bender91a} and the velocity compilation from the NASA Extragalactic Database.

To further improve the signal-to-noise ratio we folded the profiles above by taking the weighted average of bins on opposite sides of the galaxy (i.e., bins $i$ and $-i$), using the expected anti-symmetry and symmetry of the rotation and dispersion profiles, respectively. Values of the observed $V_{\rm rms}$ and their formal errors were obtained by adding the results for $V$ and $\sigma$ in quadrature.  The full kinematic profiles are shown in Figure~\ref{fig_vp}; the folded profiles are shown in Figure~\ref{f:kin}.


\subsection{The Velocity Profiles and Rotation in dE Galaxies}\label{ssec_vp}


We present the major-axis velocity profiles for our two dEs in Figure~\ref{fig_vp}.  The profiles extend to slightly more than $14'$ in each galaxy, equivalent to 7\,$r_{\rm eff}$ and 9\,$r_{\rm eff}$ in NGC~147 and NGC~185, respectively.  Previous kinematic observations for each dE have been limited to within one $r_{\rm eff}$ \citep{simien02a}.  In Figure~\ref{fig_vp}, a clear rotation signal is seen even in the individual unbinned velocity distribution of stars.  The binned major-axis velocity profiles in this figure show a well-defined rotation curve profile.  The maximum observed rotational velocity is $17\pm 2$\kms\ and $15\pm 5$\kms\ for NGC~147 and NGC~185, determined from the maximum data point in the folded rotation curves (Figure~\ref{f:kin}).  While our velocity profiles flatten at large radius, it is unlikely that, even at these large radial distances, we have reached the maximum rotation velocity in these galaxies.   The velocity dispersion profiles in both dEs decline gently with radius.  The weighted average values are $\sigma = 16\pm1$\kms\ and $24\pm1$\kms\ for NGC~147 and NGC~185, respectively.

The significant rotation velocities we measure are in fact consistent with literature results of little to no rotation.  Previous observations of NGC~147 and NGC~185 were limited to within $2'$, corresponding to the inner three data points of Figure~\ref{fig_vp}.  Fitting a simple straight line to our inner data points,  we predict a rotation velocity of merely 0.8\kms\ at this radius for NGC~185, well below the observational errors of both \citet{bender91a} and \citet{simien02a}, and fully consistent with their conclusion of no rotation.  For NGC~147, we predict a rotation velocity of 4.3\kms\ within $2'$.  This is marginally inconsistent with both \citet{simien02a} who found no rotation at this radius, and \citet{bender91a} who measured $6.5\pm1.1$\kms, but is consistent with the average of these two observations.  Our discovery of rotation at radii larger than $r_{\rm eff}$ has significant implications for studies of more distant dE galaxies. We will discuss this issue further below and in \S\,\ref{sec_disc}.

Early kinematic observations of the Local Group dEs \citep{davies83a,bender91a} established the paradigm that dE galaxies are supported by anisotropic velocity dispersions, in contrast to low-luminosity elliptical galaxies which are rotationally-supported.  Our observations show that the Local Group dEs do indeed have significant rotational support.   To demonstrate this, we compare the ratio of the observed and predicted values of the maximum rotational velocity to the average velocity dispersion using the quantity $(V_{\rm max}/\sigma)^*$.    For a rotationally-supported oblate galaxy, $(V_{\rm max}/\sigma)^*= 1$; values less than unity imply anisotropic-support.  Using the measured ellipticity of each dE (Table~1), the measured maximum rotation velocity and the average velocity dispersion values from above, we compute $(V_{\rm max}/\sigma)^* = 0.95$ for NGC~147 and $(V_{\rm max}/\sigma)^* = 0.91$ for NGC~185.  Because our observed maximum rotational velocities are lower limits on the true maximum rotation, these $(V_{\rm max}/\sigma)^*$ values are
also lower limits to the true values.   For the third dE in the Local Group, NGC~205, we measured $(V_{\rm max}/\sigma)^* = 0.3$ in \citet{geha06a}.  The original value for NGC~205 may have been somewhat larger given that the outer parts of this galaxy have been tidal disrupted by interactions with M\,31.  Thus, while the majority of kinematic support for the observed shapes of the Local Group dE galaxies comes from rotation, some anisotropic velocity dispersions are needed to fully explain the observed shapes.   This is consistent with our dynamical modeling results in \S\,\ref{ss:compare}.

\subsection{Metallicity}\label{subsec_feh}

We measure the spectroscopic metallicities ([Fe/H]) of individual RGB stars based on the \ion{Ca}{2} triplet absorption lines near $\lambda = 8500$\,\mbox{\AA} .  This method relies on an empirical calibration to Galactic globular clusters and is reliable for metallicities above [Fe/H]$> -2.5$ \citep{kirby08a}, which is the case for our member stars.  We calculate the equivalent widths (EWs) of the three \ion{Ca}{2} absorption lines using the line and continuum definitions of \citet{rutledge97a} and convert this to [Fe/H] using the empirical calibration relationship of \citet*{rutledge97b}. We determine the error on the [Fe/H] values with the Monte Carlo method described in \S\,\ref{subsec_rvel}, and add in quadrature a systematic uncertainty of $0.3$\,\mbox{\AA} as determined in \citet{simon07a}.  We restrict this analysis to stellar spectra with signal-to-noise ratio greater or equal to 5 per pixel.

The binned metallicity profiles for NGC~147 and NGC~185 are shown in the bottom panels of Figure~\ref{fig_vp}.  For NGC~147, we find an average metallicity of [Fe/H]$=-1.1\pm0.1$ and for NGC~185 we find [Fe/H]$=-1.3\pm0.1$.  This is in agreement with previous measurements from a variety of sources, including photometric and spectroscopic metallicity estimates \citep{han97a,martinez-delgado99a, davidge05a,goncalves07a}.  While the mean metallicity is well determined, there is significant internal metallicity dispersion of 0.5\,dex in each galaxy, assuming a Gaussian metallicity distribution function. Although the metallicity distribution function is asymmetric, given our errors the assumption of a Gaussian distribution is justified.  In Figure~\ref{fig_vp}, we plot both the error on the mean (barred error bars) in each radial bin and the internal metallicity dispersion (unbarred errors).  Despite our significant radial coverage, we do not find evidence for radial metallicity gradients.

\section{Dynamical Modeling} \label{sec_model}

We have constructed a set of dynamical models to provide an initial interpretation of the data.  The software was developed by \citet{vandermarel94b} for modeling of the Local Group galaxy M32 and used more recently in \citet{vandermarel07a}.  The models assume a two-integral form of the distribution function and are used to fit the binned profiles in Figure~\ref{f:kin}.  Binning results some information loss and methods do exist to model unbinned data \citep[e.g.~][]{strigari08a, chaname08a, coccato09a}, however we chose to bin the kinematic data here primarily to facilitate comparisons to previous work on these galaxies.   A full analysis of this dataset should preferably avoid spatial binning of the data, restrictive use of only the lowest-order velocity moments and allow for a more general form of the distribution function.  We intend to construct such models in a future paper using the discrete Schwarzschild method of \citet{chaname08a}.

Our present models are based on solutions of the Jeans equations of hydrostatic equilibrium with the following assumptions: (1) an oblate axisymmetric geometry with a constant axial ratio; (2) a constant mass-to-light ratio $M/L$; and (3) a two-integral distribution function of the form $f=f(E,L_z)$, where $E$ is the energy and $L_z$ the angular momentum around the symmetry axis.  Two-integral models have $\overline{v_R^2} \equiv \overline{v_z^2}$, so they can be thought of as the axisymmetric generalization of spherical isotropic models. A summary of the equations on which the models are based is provided in Section~2 of \citet{vandermarel07a}. We do not repeat this here, but focus instead on the specific application to NGC~147 and NGC~185.  Quantities for these galaxies that are relevant to the models are listed in Table~\ref{t:prop} and are discussed below.


\begin{figure*}[t]
\epsfxsize=0.7\hsize
\centerline{\epsfbox{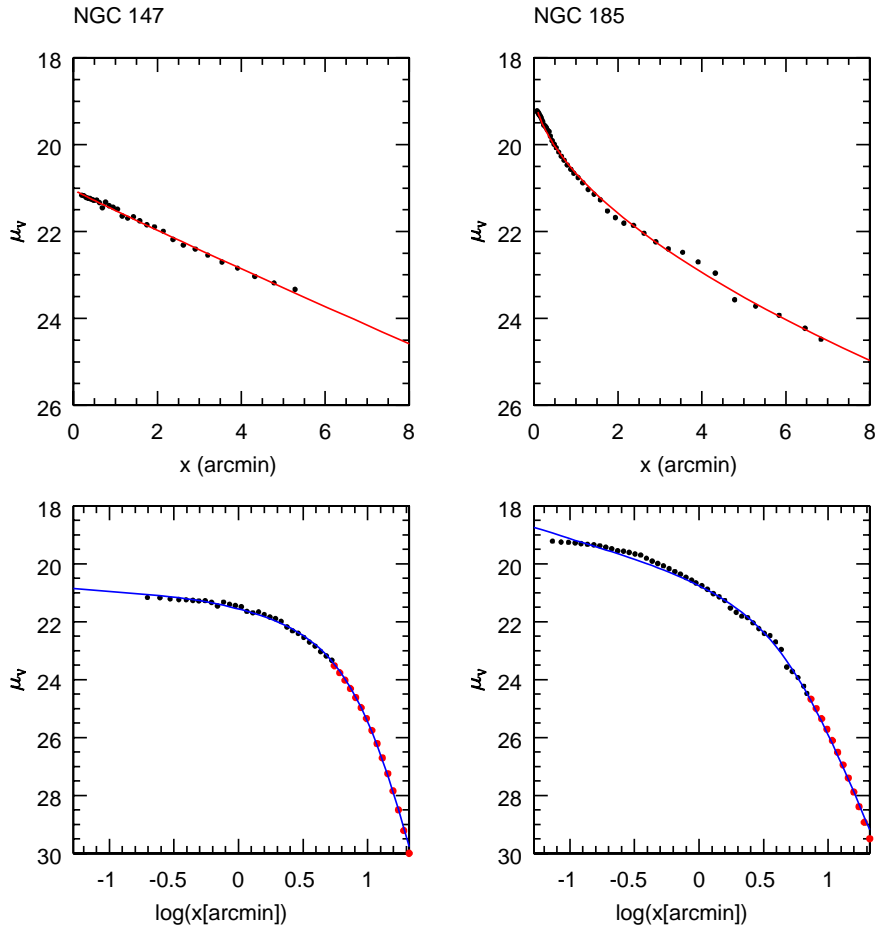}}
\figcaption{Extinction-corrected $V$-band surface brightness 
profiles in mag/arcsec$^2$ of NGC~147 ({\it left}) and NGC~185
({\it right}). Black dots show the data from \citet{kent87a}, transformed to the
$V$-band. Data inside of $1.5$ times the seeing FWHM were ignored. The
top panels have the linear major-axis radius $x$ along the abscissa,
while the bottom panels have $\log(x)$ along the abscissa. Red curves
in the top panels are the best Sersic-profile fits to the data. Red
dots in the bottom panels show the Sersic extrapolation of the data to
larger radii.  Blue curves in the bottom panels are the predictions of
the best-fitting luminosity density parameterization $j(R,z)$ in
equation~(\ref{rhodef}). Parameters of the fits are listed in
Table~\ref{t:prop}.\label{f:sb}}
\end{figure*}


\subsection{Brightness and Density Profiles} 
\label{ss:prof}

The modeling starts with the observed surface photometry of each galaxy.  While the CFHT data discussed in \S\,\ref{subsec_targets} are appropriate for measuring surface brightness profiles, an unrecoverable calibration error resulted in a significant zero-point error between
CCDs.   Instead, we use the data presented by \citet{kent87a} in the Gunn $r$-band.  We transformed the Kent major-axis data to the Johnson $V$-band using the $(B-V)$ colors from the RC3 (see Table~\ref{t:prop}) combined with the transformation equation $V-r = 0.486 (B-V) - 0.273$ \citep{Jorgensen94a}.  The resulting surface brightness profiles, corrected for extinction \citep{schlegel98a}, are shown in Figure~\ref{f:sb} and extend to $5.3'$ and $6.8'$ for NGC~147 and NGC~185, respectively. For both galaxies this is less than half the extent of our kinematical data. To extrapolate the photometry to larger radii we determined the best-fitting Sersic profiles, which are over-plotted in the top panels of Figure~\ref{f:sb}. The profiles fit the data well, with RMS residuals of $0.04$ and $0.08$\,mag  for NGC~147 and NGC~185, respectively.

We set the apparent axial ratio $q_a$ ($q_a = 1 - \epsilon$) equal to the average of the observed axial ratios in the radial range $1'-5'$ (see Table~\ref{t:prop}). Any systematic variations in axial ratio over this radial range are small in both galaxies ($\lta 0.1$; Kent 1987).  The total apparent magnitude, $V_0$, calculated using this axial ratio and our best fitting Sersic model above agree with that in the RC3 catalog to within $\sim 0.1$--$0.2$ magnitudes for each dE galaxy. This provides added confidence in the use of Sersic extrapolations at large radii.

The inclination of each dE is not well constrained by the data, however, both the predicted kinematical profiles and the inferred $M/L$ are generally insensitive to the assumed inclination \citep[e.g.\,][]{vandermarel07a,vandermarel07b}. The distribution of intrinsic axial ratios $q_t$ for elliptical galaxies is known \citep{tremblay95a}, and we assume this distribution for our dEs. We use this distribution as in \citet{vandermarel07a} to calculate the statistically most likely inclination. Since NGC~147 appears flatter on the sky than NGC~185, its most likely inclination is closer to edge-on ($i=78^{\circ}$) than that for NGC 185 ($i=64^{\circ}$). The intrinsic axial ratios $q_t$ corresponding to these assumed inclinations are given in Table~\ref{t:prop}.  We discuss the dependence on inclination in \S\,\ref{ss:ML}.
 
We parameterize the three-dimensional luminosity density as:
\begin{equation}
\label{rhodef}
  j(R,z) = j_0 (\frac{m}{b})^{\alpha} [1+(\frac{m}{b})^2]^{\delta} , \quad
           m^2 \equiv R^2 + z^2 q_t^{-2} .
\end{equation}
Here $z$ is the symmetry axis of the galaxy and $(R,\phi,z)$ are the usual cylindrical coordinates. We fit the projected surface brightness (obtained by numerical line-of-sight integration of $j$) to the combination of the Kent~(1987) data and its Sersic-profile extrapolation to $20'$. The surface brightness profile fits are shown in the bottom panels of Figure~\ref{f:sb} and the model parameters are listed in Table~\ref{t:prop}. These results are independent of the assumed inclination, although $j_0$ depends on inclination through the ratio $(q_t/q_a)$, as well as on the adopted distances. The surface brightness models fit the extrapolated profiles adequately, with RMS residuals out to $20'$ of $0.06$ and $0.10$\,mag for NGC~147 and NGC~185, respectively.


\begin{figure*}[t]
\epsfxsize=0.8\hsize
\centerline{\epsfbox{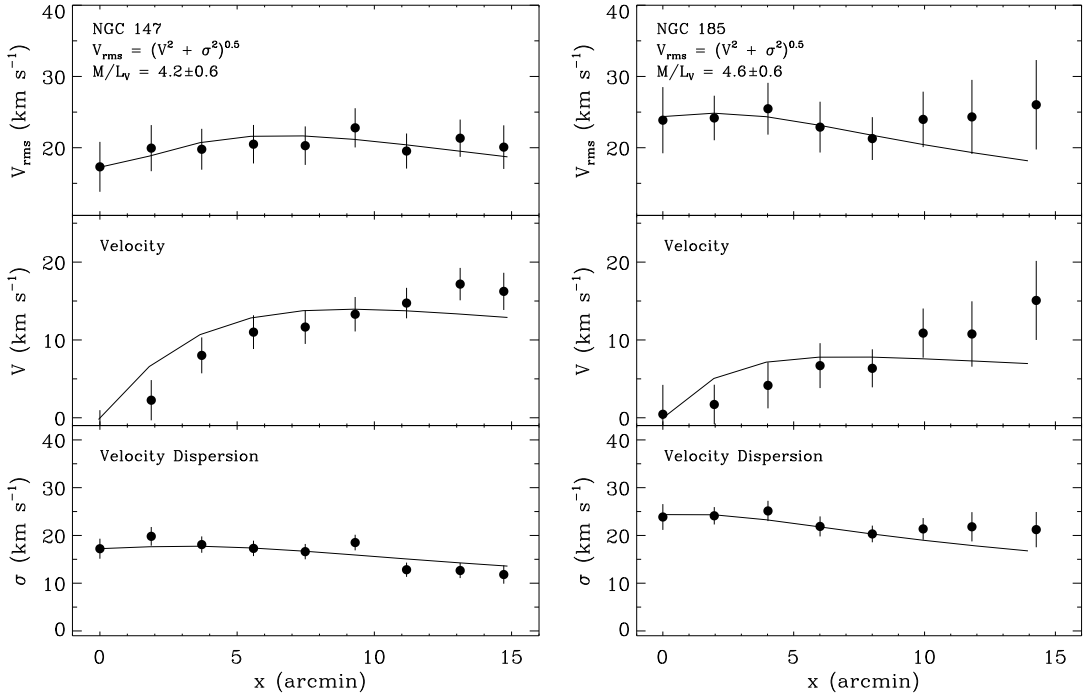}}
\vskip -0.5cm
\figcaption{Major-axis kinematics of NGC~147 ({\it left}) and 
NGC~185 ({\it right}). Black dots show the binned results extracted from our
data as described in the text. Black curves show the predictions of
the best-fitting $f(E,L_z)$ models. Parameters of the fits are listed
in Table~\ref{t:prop}. 
The top panels provide
evidence for an $M/L$ increase with radius in NGC~185, but not in NGC~147.\label{f:kin}}
\end{figure*}


\subsection{Dynamics}
\label{ss:dyn}

Given the $j(R,z)$ luminosity density, we assume a constant $M/L$ to calculate the gravitational potential. We then solve the Jeans equations to determine the intrinsic dynamical quantities, followed by a luminosity-weighted line-of-sight projection to calculate observable quantities. In particular, the models yield the quantity $V_{\rm rms} \equiv [V^2 + \sigma^2]^{1/2}$, where $V$ is the mean velocity and $\sigma$ is the velocity dispersion. The quantity $V_{\rm rms}$ scales linearly with $(M/L)^{1/2}$, but its spatial and radial dependence are otherwise uniquely determined. The quantities $V$ and $\sigma$ can also be predicted separately, but this requires the introduction of a free parameter (or function) $k$ that specifies how the second azimuthal velocity moment separates into mean and random components:
\begin{equation}
\label{satohk}
   \overline{v_{\phi}} = k [\overline{v_{\phi}^2} - \overline{v_R^2}]^{1/2} .
\end{equation}
For $k=0$ the model is non-rotating, whereas for $|k|=1$ the velocity
dispersion tensor is isotropic and the model is a so-called ``oblate
isotropic rotator''. The quantity $k$ is therefore similar to the
quantity $(V/\sigma)^*$ discussed is \S\,\ref{ssec_vp}, with the
difference that $k$ is defined locally while $(V/\sigma)^*$ is defined
in terms of globally averaged projected quantities.

\subsection{Data-Model Comparison}
\label{ss:compare}

For proper comparison to the data, the model predictions were binned
on the sky in rectangles of size $\Delta x \times \Delta y$. Here $y$
is the minor-axis direction. The quantity $\Delta y$ was taken to be
the approximate extent of the region around the major-axis over which
data was obtained. For NGC~185 we also took into account that the
observed stars are on average somewhat offset by $\delta y \approx
-1.5'$ from the major-axis (Figure~\ref{fig_dss}). No luminosity
weighting was applied in calculation of the binned model predictions,
since the observed stars are distributed approximately homogeneously
in each bin. 

The value of $M/L$ was chosen to best-fit the folded $V_{\rm rms}$ profile near the galaxy center, which in practice was chosen to be the region inside $4'$.  We determine mass-to-light ratios of $M/L_V = 4.2\pm0.6$ and $M/L_V = 4.6\pm0.6$ for NGC~147 and NGC~185, respectively.  The value of $k$ in our models was chosen to best fit the rotation curve at all available radii. The parameters of the best-fitting models are listed in Table~\ref{t:prop}. Figure~\ref{f:kin} compares the corresponding model predictions to the observed kinematical profiles.  We discuss the $M/L$ values further in \S\,\ref{ss:ML}.

For NGC~147, the predicted profile provides a
statistically acceptable match to the $V_{\rm rms}$ data ($\chi^2 = 3.8$ for 8
degrees of freedom or $\chi_{\nu}^2 = 0.5$ ).  Although the $M/L$ was chosen to only optimize
the fit near the center, the predicted profile fits the
full observed profile out to the last measured data point. Hence, even data
out to $\sim 7.3 r_{\rm eff}$ provides no evidence for an increase in
$M/L$ with radius. By contrast, for NGC~185, the predicted $V_{\rm
  rms}$ profile does {\it not} provide a statistically acceptable
match to the data. The fit is reasonable inside $4'$, since the
$M/L$ was chosen to optimize the fit there.  However, for the data
points outside of this radius the observed weighted average $V_{\rm
  rms} = 23.0 \pm 1.1$\kms. The average model prediction at these
radii is $V_{\rm rms} = 20.6$\kms. Therefore, the observations exceed
the model predictions at large radii with $2.2\sigma$
significance. The most natural interpretation of this result is that
the $M/L$ of NGC~185 is increasing with radius, although tides may also produce this affect (see \S\,\ref{subsec_bound}).

The predicted rotation curves for both galaxies provide a statistically acceptable match to the  data ($\chi^2 = 11.9$ and $7.6$, for 8 and 7 degrees of freedom, for NGC~147 and NGC~185, respectively).  However, inspection of the predicted rotation curves shows that they may be slightly too high at small radii and slightly too low at large radii. This could easily be improved by choosing $k$ to be an increasing function of radius. This would be no less arbitrary than choosing it to have a constant value as a function of radius, as we have done here.  If the model also provides a statistically acceptable fit to $V_{\rm rms}$, as is the case for NGC~147, then this guarantees that the velocity dispersion $\sigma$ is fit as well. By contrast, for NGC~185 an increase of $M/L$ with radius might be a more natural way to improve the fit to the rotation curve, given the $V_{\rm rms}$ results. The best-fitting $k$ values in Table~\ref{t:prop} indicate that while both galaxies have significant rotation rates, neither galaxy rotates fast enough to account for its flattening. Hence, the velocity dispersion tensors must be anisotropic, in agreement with our conclusion from \S\,\ref{ssec_vp}.

The fact that NGC 147 is well fit by a constant-$M/L$ two-integral model does not mean that it is not embedded in a dark halo. The profile of $V_{\rm rms}$ with radius is almost flat. So it is likely that models with a logarithmic gravitational potential can fit the data equally well. Conversely, the fact that NGC 185 is not well fit by constant $M/L$ two-integral model does not mean that it necessarily {\it must} have a dark halo. It may be possible to fit the data with a constant-$M/L$ model that has a different velocity dispersion anisotropy than assumed here.  Alternatively, our assumption of dynamical equilibrium may be incorrect for NGC 185 at large radius and we discuss the possibility of tidal interactions in \S\,\ref{subsec_bound}.  In principle it is possible to constrain the velocity dispersion anisotropy independent of the assumed gravitational potential by using information on deviations of the line-of-sight velocity distributions from Gaussians \citep[e.g.,][]{carollo95a,rix97a}.  To further explore these issues, we intend to use the discrete Schwarzschild method of Chaname \etal (2009) to construct dynamical models with three-integral distribution function, including a dark halo (i.e., $M/L$ varying with radius), which will avoid spatial binning of the data.  Such models will not necessarily yield smaller error bars or better constraints on quantities of interest (such as the mass-to-light ratio or presence of a dark halo), given the limited number of stars available to trace the kinematics. However, they will have the benefit of yielding more rigorous error bars and constraints, which include, e.g., modeling uncertainties due to the unknown velocity anisotropy and radial mass-to-light ratio profiles of these systems.


\begin{figure*}[t]
\epsfxsize=0.8\hsize
\centerline{\epsfbox{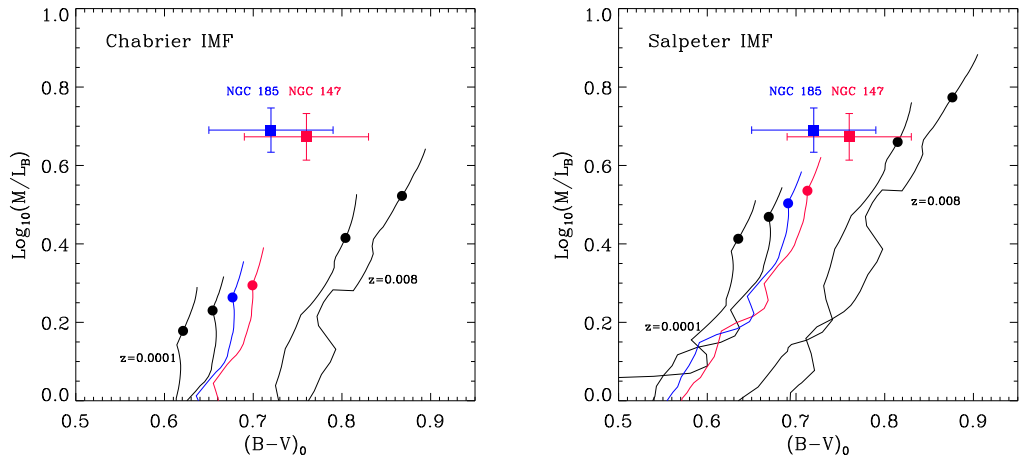}}
\figcaption{$M/L_B$ versus extinction corrected color $(B-V)_0$.
Observed values for NGC 147 (red) and NGC 185 (blue) are shown as
data points with error bars. Curves show model predictions from
Charlot \& Bruzual (2007) for the IMFs from Chabrier (2003; left) and
Salpeter (1955; right). Each curve is for a fixed metallicity $Z$ and
connects the predictions of simple stellar populations of fixed age.   Solid circles indicate ages of 10\,Gyr for each metallicity.  The predicted curves extend up to the age of the Universe \citep[13.73\,Gyr;][]{spergel07a}. Black curves are, from left to right, $Z=0.0001$,
$0.0004$, $0.004$, and $0.008$. The red and blue curves shows the interpolation of
these curves to the metallicity appropriate for NGC~147 and
NGC~185, respectively.   We note for the preferred Chabrier IMF, the inferred $M/L_B$ ratios 
of these dE galaxies cannot be explained via their stellar content alone.
\label{f:pop}}
\end{figure*}


\subsection{Mass-to-Light Ratios}
\label{ss:ML}

We compare our inferred $M/L$ ratios with stellar population models to determine whether or not non-baryonic dark matter is needed to explain our observed kinematics.  
Despite our restrictive model assumptions, our resulting $M/L$ values are likely very robust. Both \citet{vandermarel07b} and  \citet{cappellari06a} have compared the results of several dynamical modeling studies, showing that different modeling approaches produced consistent $M/L$ ratios to within a systematic accuracy of $\sim 5$\%. This is true despite the use of varying data sets (long-slit and integral field data) and modeling approaches (two- and three-integral models, and models with and without dark halos). 
In the context our models,  we have further explored the robustness of our inferred $M/L$ values by constructing models with a range of inclinations (from edge-on to inclinations corresponding to models as flat as $q_t = 0.4$). Consistent with previous studies we find that both the $M/L$ values and the predicted profiles of $V_{\rm rms}$ with radius are insensitive to the assumed inclination (well within the error bars). We also explored models with different ways of extrapolating Kent's (1987) surface brightness data to larger radii. Again, we found the effect on the $M/L$ values and the predicted profiles of $V_{\rm rms}$ with radius to be well within the error bars.

We first compare our $M/L$ values to those of \citet{derijcke06a} who modeled NGC~147 and NGC~185 using a three-integral method.  For the comparison, we transform our $M/L_V$ values to the $B$-band.  The $B$-band values were obtained from the modeled $V$-band values using the galaxy $(B-V)_0$ color (Table~3) and the solar color $(B-V)_{\odot} = 0.65$.  The de Rijcke et al.~value for NGC 147, $M/L_B = 4.0_{-2.4}^{+3.2}$, is consistent with our result.  However, their value for NGC~185, $M/L_B = 3.0_{-0.7}^{+1.0}$, is lower than our result at $1.5\sigma$ significance.  The underlying reason for this may be that NGC~185 has an increasing $M/L$ with radius, as suggested by Figure~\ref{f:kin}. The best-fitting $M/L$ is then likely to depend on the radial extent of the data, which is larger for our data set than for the integrated-light measurements of \citet{derijcke06a}.


Figure~\ref{f:pop} shows the mass-to-light ratio $M/L_B$ for each dE versus the extinction corrected color $(B-V)_0$.  We compare these measurements to the stellar population predictions of Charlot \& Bruzual (2007, priv.~comm.) which is an updated version of their \citet{bruzual03a} models. The [Fe/H] metallicities, determined in \S\,\ref{subsec_feh}, are $Z = 0.0015$ ([Fe/H] = $-1.1$) for NGC~147 and $Z = 0.0009$ ([Fe/H] = $-1.3$) for NGC~185 which assumes a solar composition of $Z_{\odot} = 0.0189$ \citep{anders89a}.  The predictions at these metallicities are over-plotted in Figure~\ref{f:pop} for both the initial mass functions (IMFs) of \citet{chabrier03a} and \citet{salpeter55a} over the range from $0.1$--$100 \Msun$. For the Chabrier IMF, the maximum $M/L_B$ ratio at the metallicity of NGC~147 and NGC~185 is attained for a population as old as the Universe \citep[13.73\,Gyr;][]{spergel07a} with values $M/L_{B, \rm max} = 2.5$ and $2.3$, respectively.  This is not as high as observed, so for the Chabrier IMF one must invoke the presence of non-baryonic dark matter to fit the observed $M/L_B$ values.  The Salpeter IMF tracks, while closer to the observations, are only marginally consistent with observations for very old stellar populations.  Stellar population studies suggest significant intermediate age stellar populations in both dEs \citep{butler05a}, thus the Salpeter IMF also implies the presence of dark matter.  The Chabrier IMF is now generally preferred over the Salpeter IMF, as the latter has too many stars at low masses to properly fit real populations \citep[e.g.,][]{bell00a, chabrier03a}.  Therefore, the high observed $M/L$ values in NGC~147 and NGC~185 must be due at least in part to the presence of dark matter.

\begin{figure}[t]
\epsfxsize=0.8\hsize
\centerline{\epsfbox{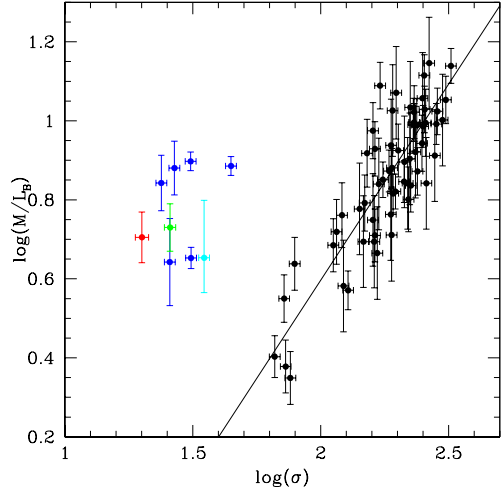}}
\figcaption{$M/L_B$ versus velocity dispersion $\sigma$. 
Observed values for NGC 147 and NGC 185 are shown in red and green
respectively. Cyan and blue data points are other dE galaxies from de
Rijcke \etal (2006) and Geha \etal (2002), respectively. Black data
points are giant ellipticals from the compilation of van der Marel \&
van Dokkum (2007b).\label{f:MLsig}}
\end{figure}

Figure~\ref{f:MLsig} shows $M/L_B$ versus the velocity dispersion
$\sigma$. For NGC 147 and NGC 185 we use along the abscissa the
weighted average $V_{\rm rms}$ inside $4'$, as listed in
Table~1.   For comparison we show in Figure~\ref{f:MLsig}
the sample of (giant) elliptical galaxies compiled by van der Marel \&
van Dokkum (2007b). The latter follow a tight linear relation of the
form $M/L_B \propto \sigma$. The dE galaxies NGC~147 and NGC~185
clearly do not follow this relation. This result is not specific to
these two dE galaxies. For comparison we plot also the results for NGC
205 from \citet{geha06a}, and for six dE galaxies in Virgo from
Geha \etal (2002). These all fall in the same region of the plot. The
same distinction between dwarf and giant ellipticals is also revealed
by plots of the fundamental plane; see e.g. the plot of $\kappa_3$
vs.~$\kappa_1$ in Geha \etal (2003). A traditional interpretation of
this is that it indicates a fundamental distinction between dwarf and
giant ellipticals. However, Graham \& Guzman (2003) have proposed more
recently that there is actually a continuous sequence between the two,
with the upturn in $M/L$ towards dE galaxies driven by the fact that
their Sersic $n$ parameter is lower than for giant ellipticals. Either
way, it is clear from Figure~\ref{f:MLsig} that dE galaxies have $M/L$
values that are similar to those of intermediate- to high-luminosity
giant ellipticals. The latter are redder and have higher metallicity
(and quite possibly older stellar populations) than NGC 147 and NGC
185. A natural explanation for the observed $M/L$ of NGC~147 and NGC~185 is that these galaxies have a significant dark matter
component, consistent with our conclusions from the comparison to
population synthesis models.


\subsection{A Bound Pair?}\label{subsec_bound}

Given our velocity and mass estimates, we revisit the question of whether or not NGC~147 and NGC~185 form a gravitationally bound pair.  The angular distance between NGC~147 and NGC~185 is $58'=10$\,kpc.  Assuming distances measured by \citet{mcconnachie05a}, NGC~147 and NGC~185 are separated by a line-of-sight distance of 59\,kpc, and thus a total physical separation of $r = 60$\,kpc.  For two point masses to be gravitationally bound, their potential energy must exceed the kinetic energy of the system \citep{davis95a, vandenbergh98a}.  This leads to the criterium $b \equiv (2GM_{\rm sys}/r \Delta v^2) > 1$ for a system to be bound, where $M_{\rm sys}$ is the combined system mass.  We measure, in \S\,\ref{sec_results}, a radial velocity difference between these two dEs of $\Delta v = 10.7 \pm 1.3$\kms.  Approximating our dE galaxies as point masses, we calculate the criteria for these two objects to be bound as $b = 1.6\pm1.1$.    This criteria suggests that, in the absence of transverse motion, NGC~147 and NGC~185 are a gravitationally bound pair,  in agreement with \citet{vandenbergh98a}.  However, the large error on this quantity, in addition to the likely non-zero transverse motions of these two satellites, leaves wide open the possibility that these are not in fact bound. 

NGC~147 and NGC~185 are satellites of  M\,31.  Our equilibrium models assume that the outer kinematics of these dEs are not affected by tidal interactions with M\,31, or each other.  This is not the case for the third dE in the Local Group, NGC~205, which is in the process of being tidal stripped at radii beyond $4.5'=1$\,kpc by M\,31.  To confirm our equilibrium assumption for NGC~147 and NGC~185, we calculate their instantaneous tidal radius in the rotating frame by again approximating these galaxies as point sources.  We assume a total mass for M\,31 between $0.8 - 1.6\times10^{12}$M$_{\odot}$ \citep{klypin02a,seigar08a}.  We determine a tidal (Jacobi) radius between 10 to 12\,kpc (greater than 25\,$r_{\rm eff}$) in each dE.  For comparison, the tidal radius of the interacting dE NGC~205 is 0.7\,kpc.   The internal kinematic profiles of NGC~147 and NGC~185 are unlikely to be currently affected by tidal interactions with M\,31, at the distances probed by the data presented here.   However, these calculations  do not preclude the possibility of past tidal interactions, either between these two galaxies or with M\,31, and it is yet possible that there are unbound stars in our kinematic sample.   The regularity and bi-symmetry of our kinematic data for NGC~147 suggest that any contamination is minimal.  In NGC~185,  the upturn of $V_{\rm rms}$ profile over our models at large radius may either be interpreted as an increasing $M/L$ ratio with radius, or possible tidal contamination.   Deeper imaging and kinematic observations are needed, particularly for NGC~185, to determine to what extent these dEs are affected by tidal interactions at radii beyond 8\,$r_{\rm eff}$.

\section{Conclusions} \label{sec_disc}

We have presented mean velocity, velocity dispersion and metallicity profiles for the Local Group dE galaxies NGC~147 and NGC~185 based on Keck/DEIMOS spectroscopic observations of 520 and 442~member RGB stars in each dE, respectively.  The profiles represent the most extensive spectroscopic radial coverage for any dE galaxy, extending to a projected distance of eight half-light radii ($14' \sim 8r_{\rm eff}$).  Contrary to previous results, we find that both dEs have significant rotation velocities which contributes to the observed flattened shapes of each galaxy.  Our two-integral dynamical modeling suggests that the observed rotation velocities cannot fully explained the observed shapes and that some anisotropic velocity dispersion is required.  Our modeling estimates $M/L$ ratios of $M/L_V = 4.2\pm0.6$ and $M/L_V = 4.6\pm0.6$ for NGC~147 and NGC~185, respectively, which is in excess of that expected from stellar populations alone.  Thus, some dark matter is required to fully explain the observe kinematics.

The mean velocity profiles of NGC~147 and NGC~185 suggest that rotation may be
far more prevalent in dE galaxies than previously assumed.  We
conclude that these Local Group dEs are rotationally-supported, however,
if placed at the distance of the Virgo or Fornax Clusters,
integrated-light observations such as those of \citet{geha03a} would
have concluded that these dEs have little to no rotation.    It is not guaranteed that the Local Group dE galaxies  had similar formation mechanisms as those residing in the much denser environments of, e.g., the Virgo or Fornax clusters. However, this is certainly plausible, given that these galaxies share similar fundamental plane properties, as defined by the photometric and kinematic properties inside $r_{\rm eff}$ (see e.g., Fig 7).  Our results then indicate that cluster dEs may also have significant rotation velocities that are manifest only at larger radii than current observations allow.  If true, this significantly modifies the observational constraints under which dEs can form.    The observations presented here open the door for formation mechanisms in which dEs are
transformed or stripped versions of gas-rich rotating progenitor galaxies.

\acknowledgments 

 We thank S.~Demers and P.~Battinelli for kindly providing their
 photometric catalogs before publication.  We also kindly thank Michael Rich for sharing Keck observing time with part of this project.  Support for this work was provided in part by NASA through Hubble Fellowship grant \#HST-HF-01233.01 awarded to ENK by the Space Telescope Science Institute, which is operated by the Association of Universities for Research in Astronomy, Inc., for NASA, under contract NAS 5-26555.




\clearpage

\begin{deluxetable}{lccccccccc}
\tabletypesize{\scriptsize}
\tablecaption{Local Group dEs at a Glance}
\tablewidth{0pt}
\tablehead{
\colhead{Name} &
\colhead{$\alpha$ (J2000)} &
\colhead{$\delta$ (J2000)} &
\colhead{Type} &
\colhead{Dist.} &
\colhead{$m_V$} &
\colhead{$A_V$}&
\colhead{$M_{V,0}$}&
\colhead{$\epsilon$}&
\colhead{$r_{\rm eff}$} \\
\colhead{}&
\colhead{(h$\,$:$\,$m$\,$:$\,$s)} &
\colhead{($^\circ\,$:$\,'\,$:$\,''$)} &
\colhead{}&
\colhead{(kpc)}&
\colhead{}&
\colhead{}&
\colhead{}&
\colhead{}&
\colhead{[$'$ (kpc)]}
}
\startdata
NGC~147  & 00:33:12 & +48:30:31 & dE5 & $675 \pm 27$ & $9.52\pm0.07$ &  0.57 & $-15.5$  & 0.44 & $2.03$ (0.40)    \\
NGC~185  & 00:38:58 & +48:20:14 & dE3 & $616 \pm 26$ & $9.18\pm0.05$ & 0.61 & $-15.7$   & 0.23 & $1.50$ (0.27)   \\
NGC~205  & 00:40:22 & +41:41:07 & dE5 & $824 \pm 27$ & $8.07\pm0.07$ & 0.21 & $-16.7$  & 0.50  & $2.17$ (0.52)  
\enddata
\tablecomments{The right ascension, declination, and morphological type are take from NASA/IPAC Extragalactic Database (NED).  We adopt the distances determined by \citet{mcconnachie05a}.  The total apparent $V$-band magnitude for each dE is taken from the RC3 catalog.  The absolute magnitude is calculated assuming an extinction values from \citet{schlegel98a}.  The ellipticity is the average value between  $1'-5'$ from \citet{kent87a}.  The effective, half-light radii are from \citet{derijcke06a}, based on near-infrared 2MASS photometry.  }
\end{deluxetable}


\begin{deluxetable}{lccrcccc}\label{table_mask}
\tabletypesize{\scriptsize}
\tablecaption{Keck/DEIMOS Multi-Slitmask Observing Parameters}
\tablewidth{0pt}
\tablehead{
\colhead{Mask} &
\colhead{$\alpha$ (J2000)} &
\colhead{$\delta$ (J2000)} &
\colhead{PA} &
\colhead{$t_{\rm exp}$} &
\colhead{\# of slits} &
\colhead{\% useful} \\
\colhead{Name}&
\colhead{(h$\,$:$\,$m$\,$:$\,$s)} &
\colhead{($^\circ\,$:$\,'\,$:$\,''$)} &
\colhead{(deg)} &
\colhead{(sec)} &
\colhead{}&
\colhead{spectra}
}
\startdata
N147\_1  & 0:32:54.2 & 48:22:48.0  &  34 & 3300 & 190 & 87\% \\
N147\_2  & 0:32:56.6 & 48:22:48.0  &  34 & 3600 & 184 & 79\% \\
N147\_3  & 0:33:35.0 & 48:37:09.5  &  34 & 3600 & 195 & 72\% \\
N147\_4  & 0:33:36.0 & 48:37:09.5  & 214 & 3600 & 183 & 82\% \\
N185\_1  & 0:39:36.0 & 48:26:06.0  &  41 & 3240 & 155 & 92\% \\
N185\_2  & 0:39:14.4 & 48:26:24.0  & 221 & 2040 & 143 & 78\% \\
N185\_3  & 0:38:39.5 & 48:14:58.2  &  41 & 3120 & 135 & 88\% \\
N185\_4  & 0:38:25.4 & 48:15:42.1  & 221 & 3060 & 139 & 42\% \\
N185\_5  & 0:39:03.6 & 48:26:24.0  &  41 & 3060 & 143 & 92\% 
\enddata
\tablecomments{Right ascension, declination, position angle and total exposure
time for each Keck/DEIMOS slitmask.  The final two columns refer to
the total number of slitlets on each mask and the percentage of those
slitlets for which a redshift was measured.  Note that the percentage of good slits
for N185\_4 is low due to a data reduction failure for 4 out of 8 CCDs.}
\end{deluxetable}


\begin{deluxetable*}{lllll}
\tabletypesize{\tiny}
\tablecaption{Observed and Model Quantities\label{t:prop}}
\tablehead{
\colhead{Row} & \colhead{Quantity} & \colhead{Units} & \colhead{NGC 147} & \colhead{NGC 185} \\ 
}
\startdata
\multicolumn{5}{l}{Photometric Properties}\\
(1) & $m_{V,o}$        & mag & $8.95\pm0.05$   & $8.57\pm0.05$   \\
(2) & $(B-V)_o$   & mag & $0.76\pm0.07$   & $0.72\pm0.07$   \\
(3) & PA$_{\rm maj}$ & degrees & 208.4   & 222.9   \\
(4) & $q_a$          &         & 0.56    & 0.77   \\ 
(5) & $i$            & degrees & 77.5    & 63.7   \\
(6) & $q_t$          &         & 0.53    & 0.70   \\ 
(7)& $n$            &         & 1.04    & 1.76   \\
(8)& $j_0(q_t/q_a)$ & $L_{\odot} pc^{-3}$ & $1.074 \times 10^{-2}$ & $3.008 \times 10^{-2}$ \\
(9)& $b$            & arcmin  & 10.72   &  5.48  \\ 
(10)& $\alpha$       &         & $-$0.940  & $-$1.439 \\
(11)& $\delta$       &         & $-$3.982  & $-$2.038 \\
(12)& $\Delta x$     & arcmin  & 1.87    & 2.00   \\ 
(13)& $\Delta y$     & arcmin  & 5.00    & 8.00   \\ 
\hline
\multicolumn{5}{l}{Dynamical Properties}\\
(14)& $V_{\rm sys}$                          & \kms  & $-$193.1 $\pm$ 0.8 & $-$203.8 $\pm$ 1.1 \\
(15)& $V_{\rm max, obs}$ & \kms          & 17 $\pm$ 2 & 15 $\pm$ 5 \\
(16)& $\langle \sigma \rangle$          &\kms & 16 $\pm$ 1 & 24 $\pm$ 1 \\

(17)& $\langle V_{\rm rms} \rangle$ & \kms & 19.1 $\pm$ 1.1 & 24.6 $\pm$ 1.2 \\
(18)& $k$                                       &           & 0.74 $\pm 0.06$ & 0.47 $\pm$ 0.08 \\
(19)& $M/L_V$        & $M_{\odot}/L_{\odot,V}$    & 4.24 $\pm 0.56$ & 4.57 $\pm$ 0.56 \\
(20)& $M/L_B$        & $M_{\odot}/L_{\odot,B}$    & 4.71 $\pm 0.69$ & 4.90 $\pm$ 0.68 \\
(21)& Mass              & $M_{\odot}$                   &  $5.6\times10^8$  & $7.2\times10^8$\\
\hline
\multicolumn{5}{l}{Metallicity}\\
(22)& [Fe/H]                      &  dex   & $-1.1\pm 0.1$     & $-1.3\pm 0.1$     \\
(23)& $\sigma_{\rm [Fe/H]}$ &  dex   & $0.4$           & $0.5$     
\enddata
\tablecomments{\tiny Quantities for the two dE galaxies, with uncertainties included where relevant.  Row~(1) lists the observed total $m_{V,o}$ magnitude from de Vaucouleurs \etal (1991; RC3), corrected for the foreground extinction listed in Table~1.  Row~(2) lists the extinction corrected color $(B-V)_0$ from RC3.  Row~(3) gives the adopted major-axis position angle, based on the outermost radius in Kent's (1987) photometry. Row~(4) lists the apparent axial ratio of the models, which is the average axial ratio in the radial range $1-5'$ in the data of Kent (1987). Row~(5) lists the statistically most likely inclination of each galaxy, based on the observed axial ratio $q_a$ and the distribution of intrinsic axial ratios $q_t$ from Tremblay \& Merritt (1995). Row~(6) lists the true axial ratio of each galaxy, based on the observed axial ratio $q_a$ and the listed inclination.  Row~(7) lists the value $n$ of the best-fitting Sersic profile, which is shown in Figure~\ref{f:sb}. Rows~(8)--(11) give the parameters of the best-fitting luminosity density parameterization $j(R,z)$ in equation~(\ref{rhodef}). Row~(12) lists the width $\Delta x$ of the bins used in binning the stellar velocities along the major-axis. Row~(13) lists the approximate extent in the minor-axis direction of the region around the major-axis over which data was obtained.  Row (14) lists the heliocentric systemic velocity $V_{\rm sys}$.  Row~(15) lists the weighted average value of $V_{\rm rms}$ inside $4'$.   Row~(16) lists the maximum observed rotational velocity, and row (17) lists the averaged observed velocity dispersion.  Rows~(18)-(20) are the mass-to-light ratio $M/L$, and the rotation parameter $k$ (see eq.~[\ref{satohk}]) inferred by fitting the models to our new kinematical data. The listed uncertainties in $M/L$ include the propagated uncertainties in the galaxy distances $D$ and the photometric uncertainties in the surface brightness profiles and color transformations. The $B$-band mass-to-light ratios were obtained from the modeled $V$-band values using the galaxy color $(B-V)_0$ and the solar color $(B-V)_{\odot} = 0.65$ (Binney \& Merrifield 1998).  The characteristic mass $M = (M/L) \times L$ of each galaxy is listed in row (21).  Row~(22) lists the average metallicity as described in \S\,\ref{subsec_feh}, row (23) lists the internal metallicity dispersion in each dE.}

\end{deluxetable*}

\begin{deluxetable}{lcccccccccc}
\tabletypesize{\scriptsize}
\tablecaption{Major-Axis Profile of NGC~147}
\tablewidth{0pt}
\tablehead{
\colhead{x} &
\colhead{$\alpha$ (J2000)} &
\colhead{$\delta$ (J2000)} &
\colhead{$V$} &
\colhead{$V_{\rm err}$} &
\colhead{$\sigma$} &
\colhead{$\sigma_{\rm err}$} &
\colhead{ [Fe/H]} &
\colhead{ [Fe/H]$_{\rm err}$} \\
\colhead{(arcmin)}&
\colhead{(h$\,$:$\,$ m$\,$:$\,$s)} &
\colhead{($^\circ\,$:$\,'\,$:$\,''$)} &
\colhead{(\kms)} &
\colhead{(\kms)} &
\colhead{(\kms)} &
\colhead{(\kms)} &
\colhead{(dex)} &
\colhead{(dex)} 
}
\startdata
$-$14.7 &   00:32:34.5 &    +48:19:25.3 & $-$15.4 &   3.3 &  14.2 &   2.4 &  $-$0.7 &   0.1 \\
$-$13.2 &   00:32:38.6 &    +48:20:35.5 & $-$18.4 &   3.2 &  16.4 &   2.3 &  $-$0.9 &   0.1 \\
$-$11.1 &   00:32:44.0 &    +48:22:08.8 & $-$16.4 &   3.2 &  16.7 &   2.3 &  $-$0.7 &   0.1 \\
$-$9.4 &   00:32:48.6 &    +48:23:27.6 & $-$11.0 &   3.5 &  21.5 &   2.5 &  $-$0.8 &   0.1 \\
 $-$7.5 &   00:32:53.4 &    +48:24:50.4 & $-$11.5 &   3.4 &  18.0 &   2.5 &  $-$0.9 &   0.1 \\
 $-$5.6 &   00:32:58.4 &    +48:26:17.2 & $-$10.0 &   2.7 &  15.5 &   2.0 &  $-$0.7 &   0.1 \\
 $-$3.7 &   00:33:03.5 &    +48:27:44.3 &  $-$6.8 &   3.8 &  21.3 &   2.8 &  $-$1.0 &   0.1 \\
 $-$1.9 &   00:33:08.3 &    +48:29:07.1 &  $-$2.6 &   3.6 &  21.1 &   2.7 &  $-$1.1 &   0.2 \\
 $-$0.1 &   00:33:12.8 &    +48:30:25.9 &  $-$1.8 &   2.8 &  17.2 &   2.1 &  $-$0.8 &   0.2 \\
  1.9 &   00:33:18.1 &    +48:31:57.0 &   1.9 &   3.7 &  18.3 &   2.9 &  $-$1.0 &   0.2 \\
  3.7 &   00:33:22.9 &    +48:33:19.8 &   8.7 &   2.9 &  16.2 &   2.2 &  $-$0.6 &   0.1 \\
  5.6 &   00:33:27.8 &    +48:34:44.8 &  12.9 &   3.7 &  20.5 &   2.7 &  $-$1.1 &   0.1 \\
  7.4 &   00:33:32.7 &    +48:36:09.4 &  11.8 &   2.8 &  15.6 &   2.1 &  $-$0.9 &   0.1 \\
  9.2 &   00:33:37.4 &    +48:37:30.4 &  14.8 &   2.8 &  16.4 &   2.1 &  $-$0.9 &   0.1 \\
 11.3 &   00:33:42.7 &    +48:39:01.4 &  13.8 &   2.5 &  10.1 &   1.9 &  $-$1.2 &   0.2 \\
 13.1 &   00:33:47.5 &    +48:40:24.2 &  16.3 &   2.8 &   9.2 &   2.2 &  $-$1.3 &   0.2 \\
 14.7 &   00:33:51.8 &    +48:41:38.8 &  17.2 &   3.5 &   7.7 &   3.2 &  $-$0.8 &   0.1 
\enddata
\tablecomments{The velocity ($V$), velocity dispersion ($\sigma$) and metallicity ([Fe/H]) as a function of the binned major-axis distance x in arcminutes away from the galaxy center.   The right ascension and declination are given  for each radial bin:  
positive radial bins on the North-East side of the galaxy, 
negative radial bins are on the South-West. }
\end{deluxetable}

\begin{deluxetable}{lcccccccccc}
\tabletypesize{\scriptsize}
\tablecaption{Major-Axis Profile of NGC~185}
\tablewidth{0pt}
\tablehead{
\colhead{$x$} &
\colhead{$\alpha$ (J2000)} &
\colhead{$\delta$ (J2000)} &
\colhead{$V$} &
\colhead{$V_{\rm err}$} &
\colhead{$\sigma$} &
\colhead{$\sigma_{\rm err}$} &
\colhead{ [Fe/H]} &
\colhead{ [Fe/H]$_{\rm err}$} \\
\colhead{(arcmin)}&
\colhead{(h$\,$:$\,$ m$\,$:$\,$s)} &
\colhead{($^\circ\,$:$\,'\,$:$\,''$)} &
\colhead{(\kms)} &
\colhead{(\kms)} &
\colhead{(\kms)} &
\colhead{(\kms)} &
\colhead{(dex)} &
\colhead{(dex)} 
}
\startdata
$-$12.3 &   00:38:22.2 &    +48:11:53.5 & $-$12.6 &   9.1 &  28.8 &   6.5 &  $-$1.4 &   0.1 \\
$-$9.9 &   00:38:29.0 &    +48:13:30.0 &  $-$9.3 &   5.4 &  24.0 &   3.9 &  $-$1.4 &   0.1 \\
$-$8.1 &   00:38:34.5 &    +48:14:46.0 &  $-$7.0 &   3.6 &  19.6 &   2.6 &  $-$0.8 &   0.1 \\
$-$5.9 &   00:38:40.6 &    +48:16:10.9 &  $-$3.6 &   6.7 &  27.6 &   4.8 &  $-$1.2 &   0.2 \\
$-$4.0 &   00:38:46.2 &    +48:17:29.4 &  $-$7.9 &   6.8 &  31.8 &   4.9 &  $-$0.9 &   0.3 \\
$-$2.0 &   00:38:52.1 &    +48:18:52.2 &  $-$6.1 &   4.5 &  26.5 &   3.2 &  $-$0.7 &   0.3 \\
$-$0.1 &   00:38:57.8 &    +48:20:12.8 &   0.5 &   3.8 &  23.9 &   2.7 &  $-$0.8 &   0.2 \\
  1.9 &   00:39:03.6 &    +48:21:33.5 &  $-$0.3 &   3.1 &  23.0 &   2.2 &  $-$0.9 &   0.1 \\
  4.0 &   00:39:09.7 &    +48:22:58.4 &   3.3 &   3.3 &  23.6 &   2.4 &  $-$0.9 &   0.1 \\
  6.0 &   00:39:15.6 &    +48:24:21.2 &   7.4 &   3.2 &  20.5 &   2.3 &  $-$1.4 &   0.1 \\
  8.0 &   00:39:21.4 &    +48:25:41.9 &   5.8 &   3.3 &  20.9 &   2.4 &  $-$1.3 &   0.1 \\
 10.0 &   00:39:27.1 &    +48:27:02.2 &  11.7 &   3.9 &  20.0 &   2.8 &  $-$1.4 &   0.1 \\
 11.9 &   00:39:32.9 &    +48:28:22.8 &  11.6 &   4.7 &  18.8 &   3.5 &  $-$1.5 &   0.1 \\
 14.4 &   00:39:40.2 &    +48:30:05.8 &  10.5 &   6.1 &  22.0 &   4.4 &  $-$1.5 &   0.1 
\enddata
\tablecomments{See comments for Table~4.}
\end{deluxetable}

\begin{deluxetable}{lccccccc}
\tabletypesize{\scriptsize}
\tablecaption{Keck/DEIMOS Velocity Measurements for NGC~147}
\tablewidth{0pt}
\tablehead{
\colhead{i} &
\colhead{Name} &
\colhead{$\alpha$ (J2000)} &
\colhead{$\delta$ (J2000)} &
\colhead{$I-$mag} &
\colhead{$(R-I)$} &
\colhead{$v$} &
\colhead{$v_{\rm err}$} \\
\colhead{}&
\colhead{}&
\colhead{(h$\,$ $\,$ m$\,$ $\,$s)} &
\colhead{($^\circ\,$ $\,'\,$ $\,''$)} &
\colhead{(mag)} &
\colhead{(mag)} &
\colhead{(\kms)} &
\colhead{(\kms)} 
}
\startdata
   1 &  11231     &     00:33:02.4 &    +48:23:16.7 &     21.3 &    0.97 &   $-$235.5 &      3.3 \\
   2 &  4456      &     00:32:23.9 &    +48:19:53.8 &     20.5 &    1.09 &   $-$235.5 &      3.3 \\
   3 &  4608      &     00:32:25.8 &    +48:18:57.2 &     20.7 &    1.14 &   $-$206.2 &      2.8 \\
   4 &  4778      &     00:32:27.6 &    +48:18:13.8 &     20.9 &    0.76 &   $-$203.4 &      4.9 \\
   5 &  4894      &     00:32:29.7 &    +48:19:03.8 &     21.1 &    0.94 &   $-$209.9 &      3.2 \\
  .. &  ..   &     .. &    .. &     .. &    .. &   .. &      ..
\enddata
\tablecomments{Velocity measurements for member stars of NGC~147.  Position, apparent $I-$band magnitude, $(R-I)$ color, heliocentric radial velocity ($v$), and velocity error ($v_{\rm err}$) for each star as determined in \S\,\ref{subsec_rvel}.  This table will be published in its entirety in the electronic edition of the {\it Astrophysical Journal}.  } 
\end{deluxetable}

\begin{deluxetable}{lccccccc}
\tabletypesize{\scriptsize}
\tablecaption{Keck/DEIMOS Velocity Measurements for NGC~185}
\tablewidth{0pt}
\tablehead{
\colhead{i} &
\colhead{Name} &
\colhead{$\alpha$ (J2000)} &
\colhead{$\delta$ (J2000)} &
\colhead{$I-$mag} &
\colhead{$(R-I)$} &
\colhead{$v$} &
\colhead{$v_{\rm err}$} \\
\colhead{}&
\colhead{}&
\colhead{(h$\,$ $\,$ m$\,$ $\,$s)} &
\colhead{($^\circ\,$ $\,'\,$ $\,''$)} &
\colhead{(mag)} &
\colhead{(mag)} &
\colhead{(\kms)} &
\colhead{(\kms)} 
}
\startdata
 1 &  11396     &     00:38:58.6 &    +48:21:52.3 &     20.5 &    0.66 &   $-$203.3 &      3.8 \\
   2 &  11487     &     00:38:58.9 &    +48:23:33.3 &     20.9 &    1.18 &   $-$202.3 &      4.4 \\
   3 &  11515     &     00:38:59.0 &    +48:22:03.4 &     20.8 &    1.01 &   $-$190.0 &      5.9 \\
   4 &  11522     &     00:38:59.0 &    +48:22:34.5 &     21.0 &    0.69 &   $-$215.6 &      4.1 \\
   5 &  12017     &     00:39:00.2 &    +48:22:09.0 &     20.1 &    0.96 &   $-$175.7 &      2.8 \\
   .. &  ..   &     .. &    .. &     .. &    .. &   .. &      ..
\enddata
\tablecomments{Same as Table~6, but for NGC~185.}
\end{deluxetable}


\clearpage


\end{document}